\documentclass[onecolumn,preprintnumbers,amsmath,amssymb,nofootinbib]{revtex4}

\usepackage[dvips]{color}
\usepackage{graphicx}
\usepackage{epsfig}
\usepackage{subfig}

\begin{document}

\title{SNe Ia Tests of Quintessence Tracker Cosmology in an Anisotropic Background}

\author{W. Miranda.$^{1,2}$\footnote{welber.miranda@ifba.edu.br}, S. Carneiro$^{2}$\footnote{saulo.carneiro@pq.cnpq.br} and C. Pigozzo$^{2}$\footnote{cpigozzo@ufba.br}}

\affiliation{$^{1}$Instituto Federal da Bahia, Paulo Afonso, BA, Brazil\\ $^{2}$Instituto de  F\'{\i}sica, Universidade Federal da Bahia, Salvador, Bahia, Brazil}

\date{\today}

\begin{abstract}
We investigate the observational effects of a quintessence model in an anisotropic spacetime. The anisotropic metric is a non-rotating particular case of a generalized G\"odel's metric and is classified as Bianchi III. This metric is an exact solution of the Einstein-Klein-Gordon field equations with an anisotropic scalar field $\psi$, which is responsible for the anisotropy of the spacetime geometry. We test the model against observations of type Ia supernovae, analyzing the SDSS dataset calibrated with the MLCS2k2 fitter, and the results are compared to standard quintessence models with Ratra-Peebles potentials. We obtain a good agreement with observations, with best values for the matter and curvature density parameters $\Omega_M = 0.29$ and $\Omega_k= 0.01$ respectively. We conclude that present SNe Ia observations cannot, alone, distinguish a possible anisotropic axis in the cosmos.
\end{abstract}

\maketitle

\section{Introduction}

Einstein's theory of relativity establishes a two-way relation between geometry and the dynamics of the energy-matter content of the universe. Within this context, a remarkable feature of our universe is that its observed dynamics demands that most of its energy originates of some peculiar fluid entity that is not gravitationally attractive. This so called dark energy has a repulsive behavior and is responsible for the increase in the expansion rate of the universe. If we concern about the dynamics of the cosmos, the dark energy content plays a major role in its evolution, representing about $3/4$ of the critical density. The evidence of its existence has been strongly corroborated by distance measurements of type Ia supernovae \cite{Perlmutter99}. The evidence accrues with measurements of the Cosmic Microwave Background (CMB) anisotropies \cite{Bernadis99,Spergel03} and clustering estimates \cite{Calberg96}. The description of dark energy has been historically modeled by the simplest of the alternatives, the Einstein's cosmological constant. The cosmological constant $\Lambda$ has great success in best-fitting different observational data with high accuracy, although there is not a robust microphysics to clarify its origin. All the same, the description by a cosmological constant is sieged with some pathologies \cite{Weinberg89}. It has the theoretical inconvenient of requiring a fine-tuning of initial conditions, adjusting $\Lambda$ to the value measured today. It is also difficult to explain why the dark energy density has initiated to dominate the cosmic dynamics so close to our era, fact known as the coincidence problem.

Different dark energy candidates can model the accelerated expansion. Among the most popular possibilities are scalar field quintessence models \cite{Steinhardt:1999nw,Zlatev98}, decaying vacuum theories \cite{Freese:1986dd,Borges:2005qs}, $f[R]$ theories \cite{Sotiriou:2008rp} and even unusual anisotropic $f[T]$ \cite{Sharif:2014fqa}. The description by means of an energy component with dynamical evolution may elucidate, with a more solid theoretical background, how the dark energy has evolved to dominate the dynamics of the universe. The quintessence tracker models may also alleviate the problem of cosmic coincidence \cite{Zlatev98} once the present dark energy density can be obtained from a wide range of initial conditions. Here we will choose a scalar field model with the Ratra-Peebles potential \cite{RatraPeebles88}. The dynamics of this type of quintessence depends on the background, and it naturally evolves to an asymptotic $\Lambda$-like behavior in the present era.

On the other hand, some observational tests need to be performed in order to analyse the possibility that our universe may not be perfectly isotropic \cite{Hansen:2004vq,Akrami,Antoniou:2010gw}. There are indications of a possible preferred direction,  sometimes called the axis of evil \cite{Kate05}. Since many observations does indicate a high degree of isotropy, this possible anisotropy must be very small in order to maintain a good agreement with those observations (see, for example, \cite{Tomi,Saridakis}). In this paper we study a Bianchi III anisotropic metric, here named RTKO (Rebou\c cas-Tiomno-Korotkii-Obukhov) metric after \cite{Reboucas83,Korotkii91} (see also \cite{Obukhov2000,Carneiromarugan01,thiago}). This spacetime is shear free, homogeneous and has conformal expansion, which guarantees the isotropy of CMB. In order to account for the anisotropy of this background, the cosmic fluid is modeled by an anisotropic scalar field $\psi$, that will be responsible for an additional pressure in the preferred direction. Choosing $\psi$ properly \cite{Carneiromarugan01,thiago,Menezes13} we can find an exact solution for the Einstein and Klein-Gordon equations.

The main goal of this work is to study the observational viability, via SNe Ia distance measurements, of quintessence models with Ratra-Peebles potential in RTKO curved spacetime. The paper is organized as follows. In Sec. II we revise the anisotropic RTKO metric and some of its properties. In Sec. III we take a detailed look at the corresponding cosmological model with dark energy described by a tracker quintessence model. In Sec. IV we analyze the model observational viability, using SNe Ia observations. We finish by summarizing our conclusions and results in Sec. V. In this paper we use $8\pi G = c = \hbar = 1$.

\section{RTKO metric}

RTKO metrics are a special case of G\"odel's metrics, which include rotation and are generally classified as Bianchi III. In Cartesian coordinates the line element is
\begin{equation}\label{RTKO_metric}
ds^2=a^2(\eta) \left[(d\eta+le^{x}dy)^2-dx^2-e^{2x}dy^2-dz^2 \right],
\end{equation}
where $\eta$ is the conformal time defined by $d\eta = dt/a(\eta)$, and $a(\eta)$ is the scale factor.  Although it can admit rotation for $0 < l < 1$, this rotation parameter will be set to zero. Our metric then reduces to
\begin{equation}\label{RTKO_metric2}
ds^2=a^2(\eta) \left[d\eta^2-dx^2-e^{2x}dy^2-dz^2 \right].
\end{equation}
It possesses the Killing vectors $\xi_1  = \partial_x-y\partial_y$, $\xi_2  = \partial_y$, $\xi_3  = \partial_z$. It has also an additional conformal Killing vector $\xi^{\mu}_C = \delta^{\mu}_0$, which guarantees the isotropy of CMB on this background \cite{Carneiromarugan03}.
This spacetime is the product of the real line and a hyperbolic manifold, $R \times H^2$. In cylindrical coordinates $x^{\mu}=(\eta,r,\phi,z)$, it is written as
\begin{equation}\label{RTKO_metric_cilindric}
ds^2=a^2(\eta)(d\eta^2-dr^2-\sinh^2r d\varphi^2-dz^2).
\end{equation}
The Einstein's equations for this metric are
\begin{eqnarray}
T_{0}^{0}a^4 &=& 3a'^2-a^2, \label{Einst.1} \\T_{1}^{1}a^4 &=& T_{2}^{2}a^4=2aa''-a'^2, \label{Einst.2} \\T_{3}^{3}a^2 &=& T_{1}^{1}a^2-1, \label{Einst.3}
\end{eqnarray}
with $T_{\nu}^{\mu}= 0$ for $\mu \neq \nu$. The prime denotes derivative with respect to the conformal time. This leads to an energy density $\rho a^4  = 3a'^2-a^2$, and to pressures $p_1 a^4  = p_2 a^4 = -2aa''+a'^2$ and  $p_3 a^2 = p_2 a^2 - 1$. A simple solution can be obtained by adding an anisotropic scalar field $\psi$ that accounts for the extra pressure in $z$ direction \cite{Carneiromarugan01}, suitably choosing $\psi$ as
\begin{equation}\label{psi_z}
\psi(z)=Cz.
\end{equation}
The Klein-Gordon equation,
\begin{equation}\label{KGE}
\psi_{;\mu\nu}=\frac{1}{\sqrt{-g}}(\sqrt{-g}\psi_{,\mu}g^{\mu\nu})_{,\nu} = 0,
\end{equation}
is simultaneously satisfied.
For the scalar field we have $T_{00}^{(\psi)}=C^2/2a^2$, $T_{11}^{(\psi)}=T_{22}^{(\psi)}=-T_{33}^{(\psi)}=C^2/2a^2$. Now we define $\rho= \bar{\rho}+\rho^{(\psi)}$ and $p = \bar{p}+ p^{(\psi)}$, to obtain
\begin{equation}\label{p1}
\bar{p}_3 a^2+p_3^{(\psi)} a^2=\bar{p}_1 a^2+p_1^{(\psi)} a^2+1,
\end{equation}
\begin{equation}\label{p2}
\bar{p}_3 a^2 = \bar{p}_1 a^2-C^2+1.
\end{equation}
If we take $C^2=1$ we have
\begin{equation}\label{pressures}
\bar{p}_1=  \bar{p}_2 =  \bar{p}_3,
\end{equation}
where now $\bar{\rho}$ and $\bar{p}$ refer respectively to the energy density and pressure of the cosmic fluid isotropic components. In this way, the Einstein's equations acquire the same form as in a FLRW open model with curvature $k = -1/2$,
\begin{eqnarray}
\bar{\rho}a^4&=&3a'^2- \frac{3}{2} a^2, \label{rho_bar}\\
\bar{p}a^4&=&a'^2-2aa''+\frac{a^2}{2}. \label{p_bar}
\end{eqnarray}
If we divide (\ref{rho_bar}) by $3a^4$, we obtain the Friedmann equation for the RTKO metric,
\begin{equation}\label{eq_hubble}
H^2 =\frac{\bar{\rho}}{3}+\frac{1}{2a^2},
\end{equation}
where $H (\eta) \equiv a'/a^2$. Now we are ready to include a dark energy component, which will account for the accelerated expansion in this model.

\section{Quintessence Tracker Model}

We will model the accelerated expantion with an isotropic and homogeneous scalar field $\phi$, which is a function of time. Our analysis will be restricted to the tracker potential of Ratra \& Peebles \cite{RatraPeebles88}. The action for this field, considering a minimal coupling to gravity, is given by
\begin{equation}\label{action}
S=\int d^4 x \sqrt{-g}  \mathcal{L} (\phi,\partial_{\mu} \phi),
\end{equation}
where the Lagrangian density is
\begin{equation}\label{lagrangian}
\mathcal{L}= \frac{1}{2} \partial_{\mu} \phi \partial^{\mu} \phi-V(\phi).
\end{equation}
The first term in (\ref{lagrangian}) corresponds to the field kinetic energy, and $V(\phi)$ is the field potencial energy. The energy density and pressure will then be
\begin{equation}\label{energypotential}
\rho_{\phi}  =  \frac{1}{2} \dot{\phi}^2+ V(\phi),
\end{equation}
\begin{equation}\label{pressurepotential}
p_{\phi}  =  \frac{1}{2} \dot{\phi}^2 - V(\phi),
\end{equation}
From the energy conservation equation we have, for the quintessence field,\begin{equation}\label{energy_consfield}
\ddot{\phi}+3H\dot{\phi}+V'(\phi)=0,
\end{equation}
where the dot and the prime denote derivatives with respect to conformal time and to $\phi$, respectively.
The potential on which we base our analysis is defined by
\begin{equation}\label{RP_potencial}
V(\phi)= \frac{M^{4+\alpha}}{\phi^{\alpha}},
\end{equation}
where $M$ is a positive constant and the parameter $\alpha$ is fixed to $\alpha = 1$ or $\alpha =2$. The Ratra-Peebles belongs to a class of potentials which exhibit a tracker behavior \cite{RatraPeebles88,Steinhardt:1999nw}, making them very suitable quintessence models, once a very large number of initial conditions leads to the same late-time behavior in the quintessence dominant era. From Einstein's equations with the RTKO metric we can build a cosmological model with radiation, matter and quintessence field. The Friedman equation gets the form
\begin{equation}\label{Friedman2}
3H^2= \rho_m+ \rho_r+ \rho_{\phi}-\frac{3k}{a^2},
\end{equation}
where $\rho_{\phi}$ is defined by (\ref{energypotential}), while $\rho_m$ and $\rho_r$ are the densities of non-relativistic matter and radiation, respectively. Now we can write
\begin{equation}\label{Friedman3}
3H^2= \rho_{r,0} \left(\frac{a_0}{a}\right)^4 + \rho_{m,0} \left(\frac{a_0}{a}\right)^3+  \frac{1}{2}\dot{\phi}^2 + V(\phi)-\frac{3k}{a^2},
\end{equation}
where $a_0$ is the present scale factor. This equation can be rewritten as
\begin{equation}\label{Friedman4}
\frac{H^2}{H_{0}^2} = \Omega_{r,0} \left(\frac{a_0}{a}\right)^4 + \Omega_{m,0} \left(\frac{a_0}{a}\right)^3+  \Omega_{\phi}+\Omega_{k,0} \left(\frac{a_0}{a}\right)^2,
\end{equation}
where $\Omega_i$ ($i = m, r, \phi$) are the relative density parameters.

\begin{figure}\includegraphics[height=5.2cm]{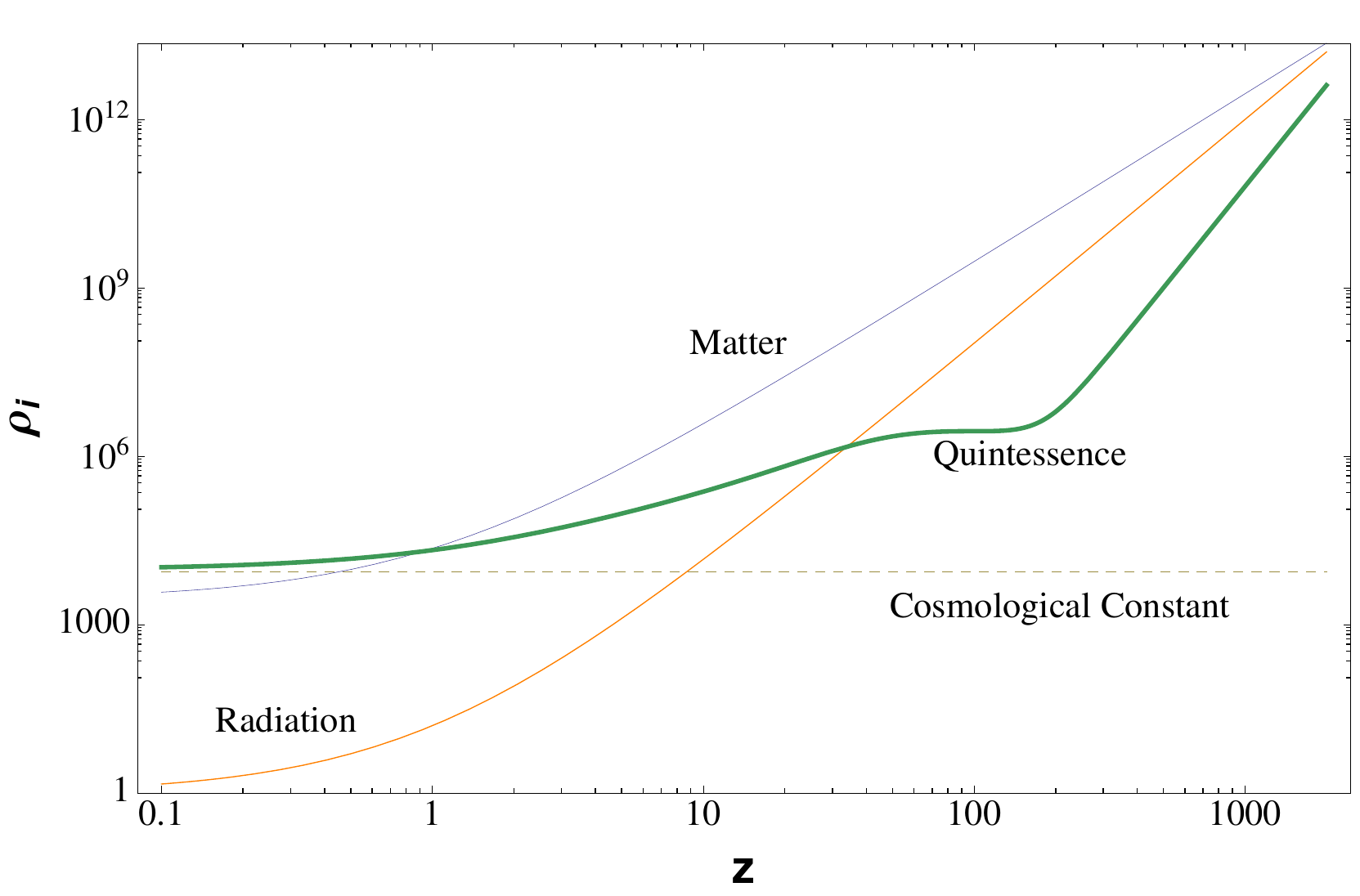}\hspace{1.2cm} \includegraphics[height=5cm]{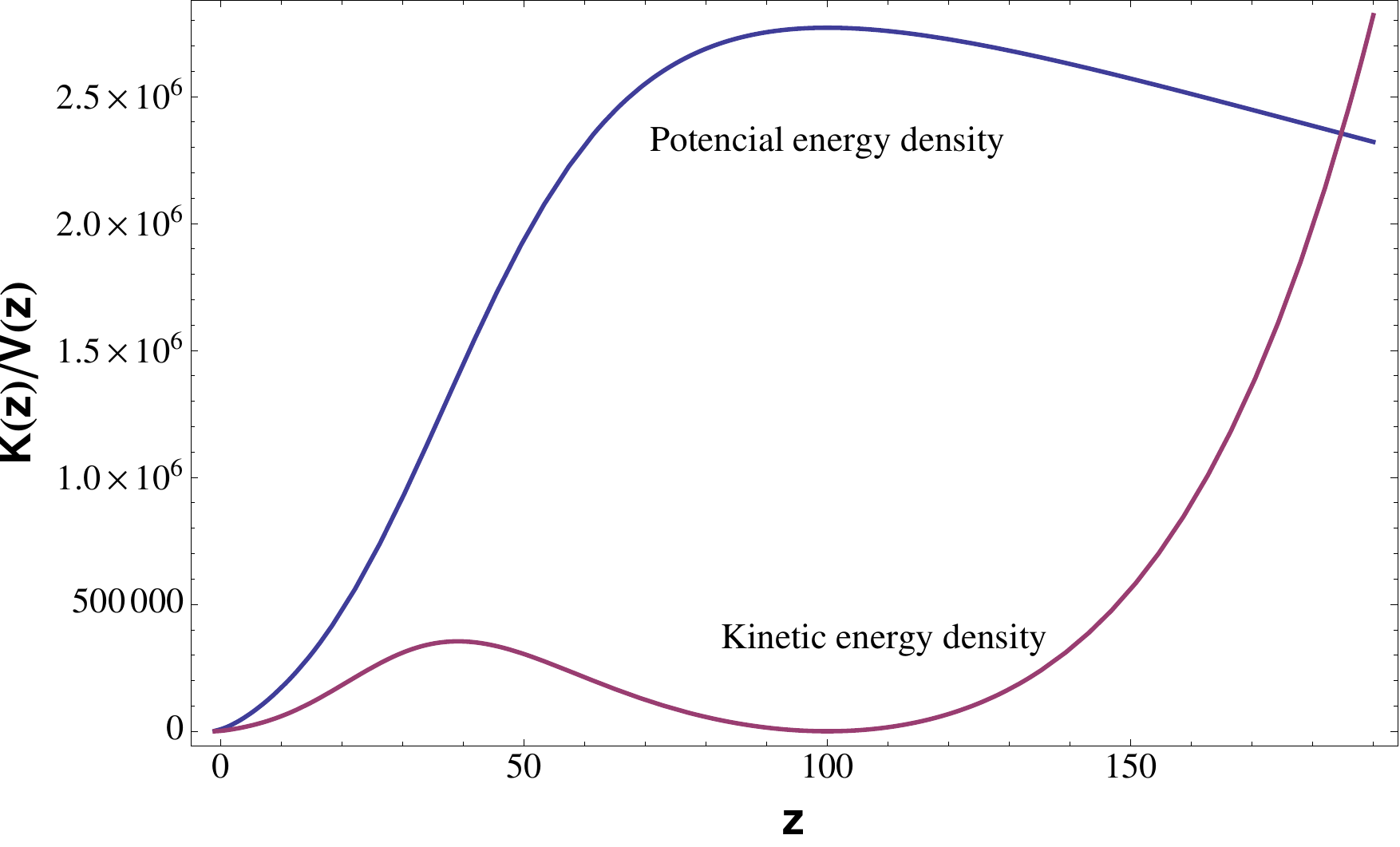}\caption{\small{Energy densities $\rho_{i}$ of the different cosmic components and the cosmological constant for reference (left) and the kinetic and potential energies of the quintessence field (right).}}
\label{fig1}
\end{figure}

\begin{figure}\includegraphics[height=5.2cm]{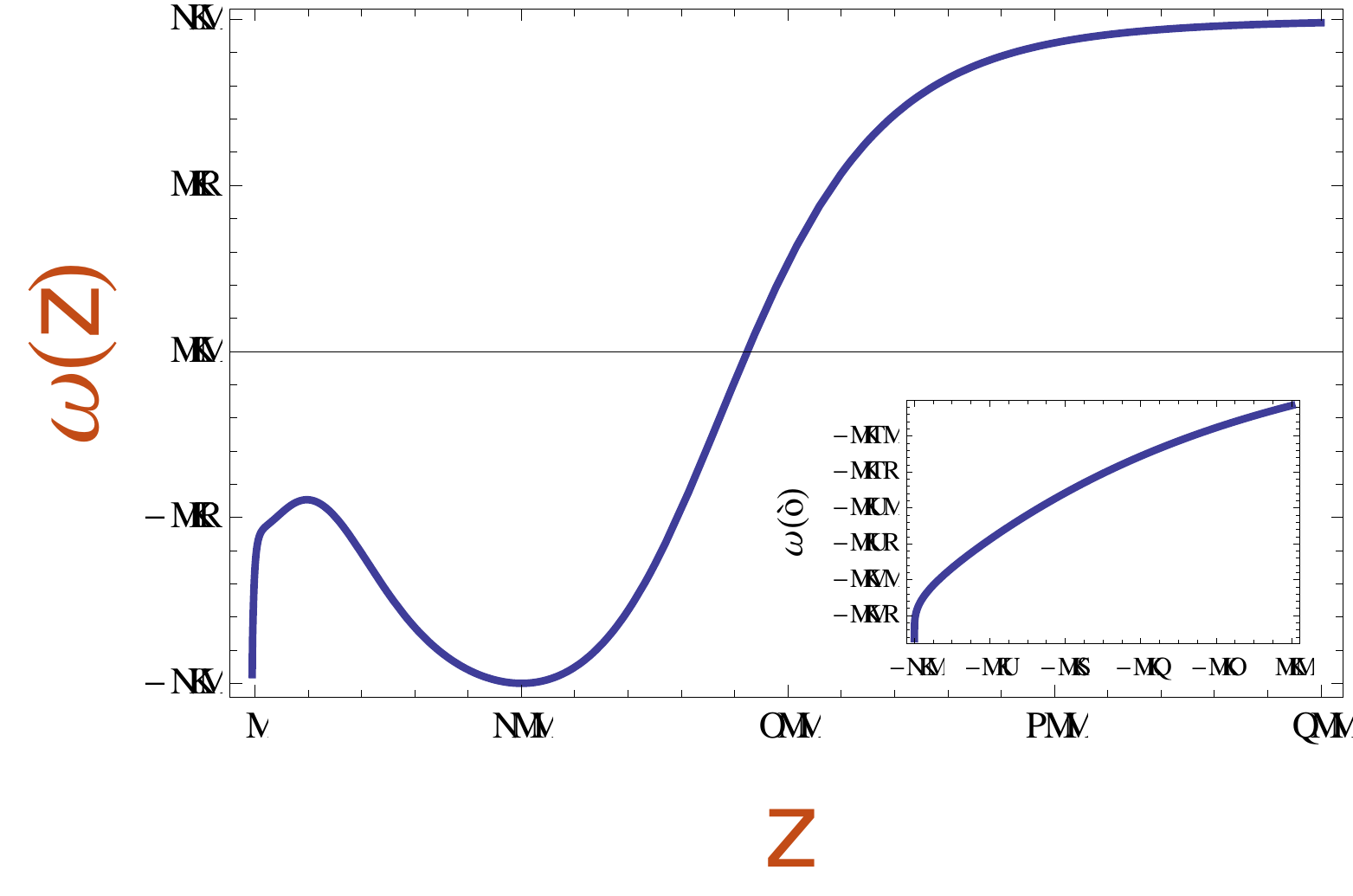}\hspace{1.cm} \includegraphics[height=5.25cm]{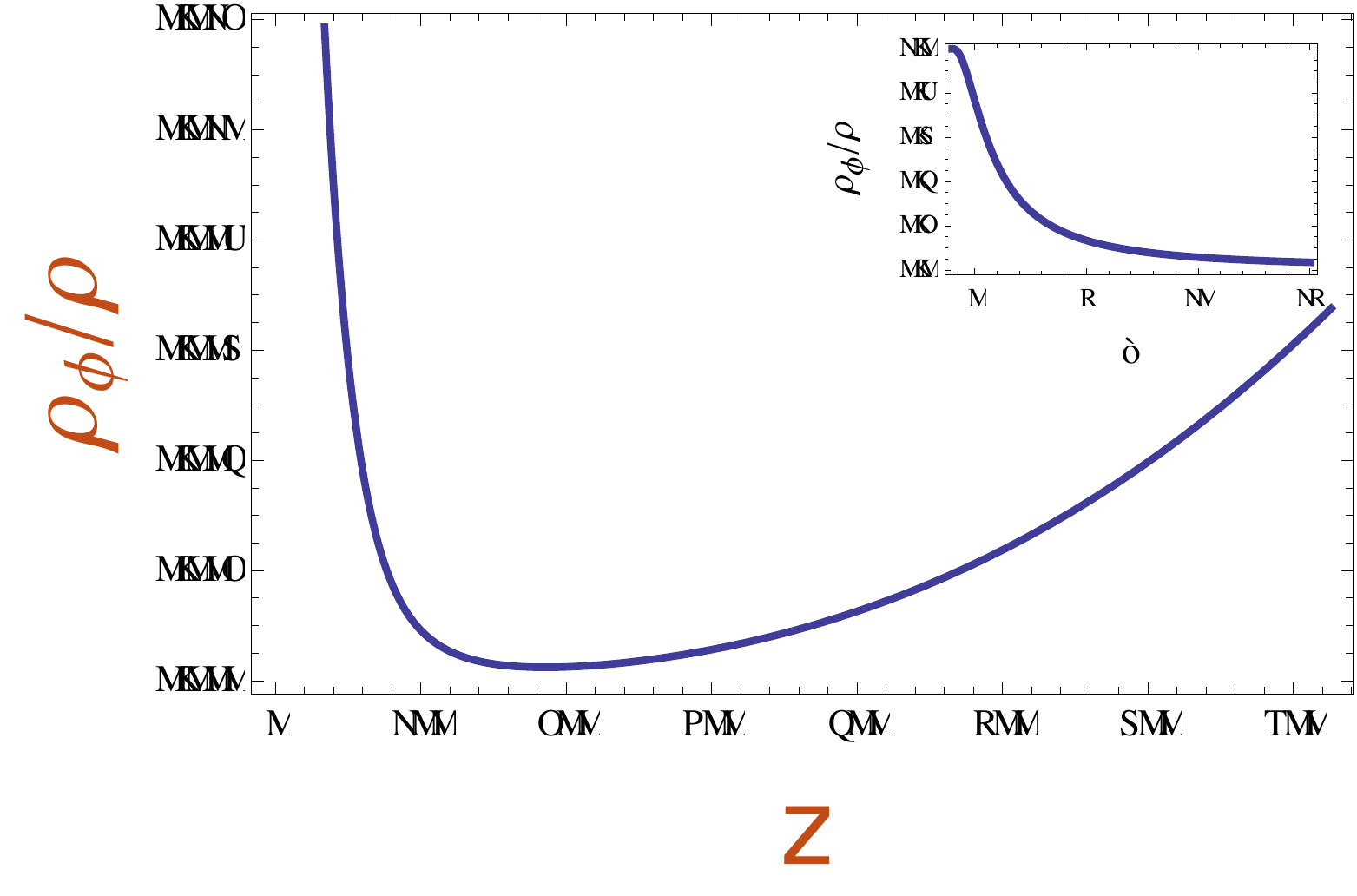}\caption{\small{Equation-of-state parameter (left) and relative energy density (right) of the quintessence component.}}
\label{fig2}
\end{figure}

As shown in figure \ref{fig1}, in the late times of cosmic evolution the quintessence scalar field dominates the dynamics of the Universe. Once radiation, matter density and curvature density can be neglected in comparison to the dark energy density, the Friedmann equation (\ref{Friedman4}) reduces to the simple form
\begin{equation}\label{Fried_field}	
H^2=\frac{\rho_{\phi}}{3},
\end{equation}
which can be put in (\ref{energy_consfield}) to give
\begin{equation}\label{eq_mov}
\ddot{\phi} +  \sqrt{3\rho_{\phi}}\dot{\phi}- \alpha \frac{M^{4+\alpha}}{\phi^{\alpha+1}} = 0.
\end{equation}
In this case the dissipation term $-\sqrt{3\rho_{\phi}}\dot{\phi}$ gradually makes the kinetic energy $\frac{1}{2}\dot{\phi}^2$ smaller than the potential energy $V(\phi)$. Hence, we can use the slow-roll approximation, doing $\dot{\phi}^2$ approximately constant, once $\ddot{\phi}$ becomes much smaller than $V(\phi)$. From $\rho_{\phi}  \approx V(\phi)$ we have
\begin{equation}\label{eq_mov2}
\sqrt{3\frac{M^{4+\alpha}}{\phi^{\alpha+1}}}\dot{\phi} = \alpha\frac{M^{4+\alpha}}{\phi^{\alpha+1}},
\end{equation}
whose solution is
\begin{equation}\label{eq_mov2_solution}
\phi(t)= M \left[ \frac{\alpha(2+\alpha/2)^2 t}{\sqrt{3}} \right] ^{\frac{1}{2+\alpha/2}} + C^1,
\end{equation}
where $C^1$ is an integration constant that can be absorbed in the definition of $t$. With the help of (\ref{energypotential}), (\ref{RP_potencial}) and (\ref{eq_mov2_solution}), the quintessence
energy density becomes
\begin{equation}\label{rho_phi_t}
\rho_{\phi}(t)  \sim t^{\frac{-\alpha}{2+\alpha/2}},
\end{equation}
and the Hubble function is given by
\begin{equation}\label{Hubble2}
H = \frac{\dot{a}}{a}  \sim t^{\frac{-\alpha}{2(2+\alpha/2)}}.
\end{equation}
We can see that the scalar field scales logarithmically as $\ln a(t) \sim t^{\frac{2}{2+\alpha/2}}$. In the case of a cosmological constant we would have $\ln a(t)  \sim t$. In figure \ref{fig2} we show the scalar field equation-of-state parameter (defined as $\omega=p/\rho$) and its relative energy density. In the insets, we can see that $\omega \rightarrow -1$ and $\Omega_{\phi} \rightarrow 1$ in the asymptotic future $z = -1$. This means that this solution asymptotically approaches a de Sitter universe, with $\rho = V=$ constant at $z=-1$, a behavior that can also be seen in the left panel of figure \ref{fig1}. In recent eras the field is still rolling with non-zero kinetic energy, a fact that may be useful to differentiate it from a constant $\Lambda$.

In terms of the redshift $z = -1+ \frac{a_0}{a}$, we can rewrite equation (\ref{Friedman4}) as
\begin{equation}\label{Hubble_z}
\frac{H(z)}{H_0} =[\Omega_{r,0} (1+z)^4+\Omega_{m,0} (1+z)^3 + \Omega_{\phi} + \Omega_k (1+z)^2 ]^{\frac{1}{2}}.
\end{equation}
The quintessence energy density parameter depends on $M$ and takes the form
\begin{equation}\label{density_phi}
\Omega_{\phi} (z) =  \frac{ \frac{1}{2} (1+z)^2 H^2{\phi '}^2}{3H_{0}^2}+ \frac{M^{4+\alpha}}{3H_{0}^2 \phi^\alpha} .
\end{equation}
If we define the function
\begin{equation}\label{E_z}	
E(z) \equiv \frac{H(z)}{H_0},
\end{equation}
we have
\begin{equation}\label{E_z3}
E(z) =  \left[ \frac{\Omega_{r,0} (1+z)^4+ \Omega_{m,0} (1+z)^3 + \frac{M^{4+\alpha}}{3\phi^{\alpha}} + \Omega_k (1+z)^2}{1- \frac{(1+z)^2 \phi'^2}{6}} \right] ^{\frac{1}{2}},
\end{equation}
where now the prime denotes derivative with respect to $z$. On the other hand, dividing (\ref{energy_consfield}) by $H_0^2$ and writing in terms of $z$ we get
\begin{equation}\label{E_z4}
E^2(z)(1+z)^2 \phi''(z)^2 + [E(z)E'(z)(1+z)^2-2E^2(z)(1+z)] \phi'(z) - \alpha  \frac{M^{4+\alpha}}{\phi(z)^{\alpha+1}}=0.
\end{equation}
Equations (\ref{E_z3}) and (\ref{E_z4}) are coupled non-linear differential-algebraic equations that governs the evolution of our anisotropic model with quintessence dark energy. To unveil its dynamics, these equations should be integrated numerically to obtain the independent functions $E(z)$ and $\phi(z)$.

\section{SN\lowercase{e} I\lowercase{a} Tests}

For a given redshift, owing to the anisotropy of the RTKO metric, a small change in the line of sight angle $\theta$ may cause a sensitive change in the corresponding distance. In this case it is more convenient to use the angular-diameter distance defined as \cite{Menezes13}
\begin{equation}\label{dA2}	
d_{2A} = \frac{a_0\chi}{1+z} \left[\frac {\sinh (\chi \sin \theta)}{\chi \sin \theta} \right]^{\frac{1}{2}},
\end{equation}
with $0 < \theta < \pi$. If we also define $Z(z) \equiv \int_0^z\frac{dz'}{E(z')}$, we can rewrite the angular-diameter distance for the RTKO metric as
\begin{equation}\label{dA2_2}
d_{2A} = \frac{Z(z)}{(1+z)H_0} \left\{ \frac{ \sinh [|\Omega_{k,0}|^{\frac{1}{2}} Z(z) \sin \theta]}{|\Omega_{k,0}|^{\frac{1}{2}} Z(z) \sin \theta} \right\}^{\frac{1}{2}}.
\end{equation}
The distance modulus, $\mu \equiv 5\log_{10} \left(\frac{d_L}{1Mpc}\right)+25$, will have the form
\begin{equation}\label{mu}
\mu = 5 \log_{10} \left[\frac{(1+z)Z(z)}{H_0} \left\{  \frac{\sinh[|\Omega_{k,0}|^{\frac{1}{2}} Z(z) \sin \theta]}{|\Omega_{k,0}|^{\frac{1}{2}} Z(z) \sin \theta} \right\}^{\frac{1}{2}} \right]+25.
\end{equation}
In the limit $\theta \rightarrow 0$, the anisotropy fades and the distance modulus reduces to that of the spatially flat FLRW model.

\begin{figure}[t]
\includegraphics[height=5cm]{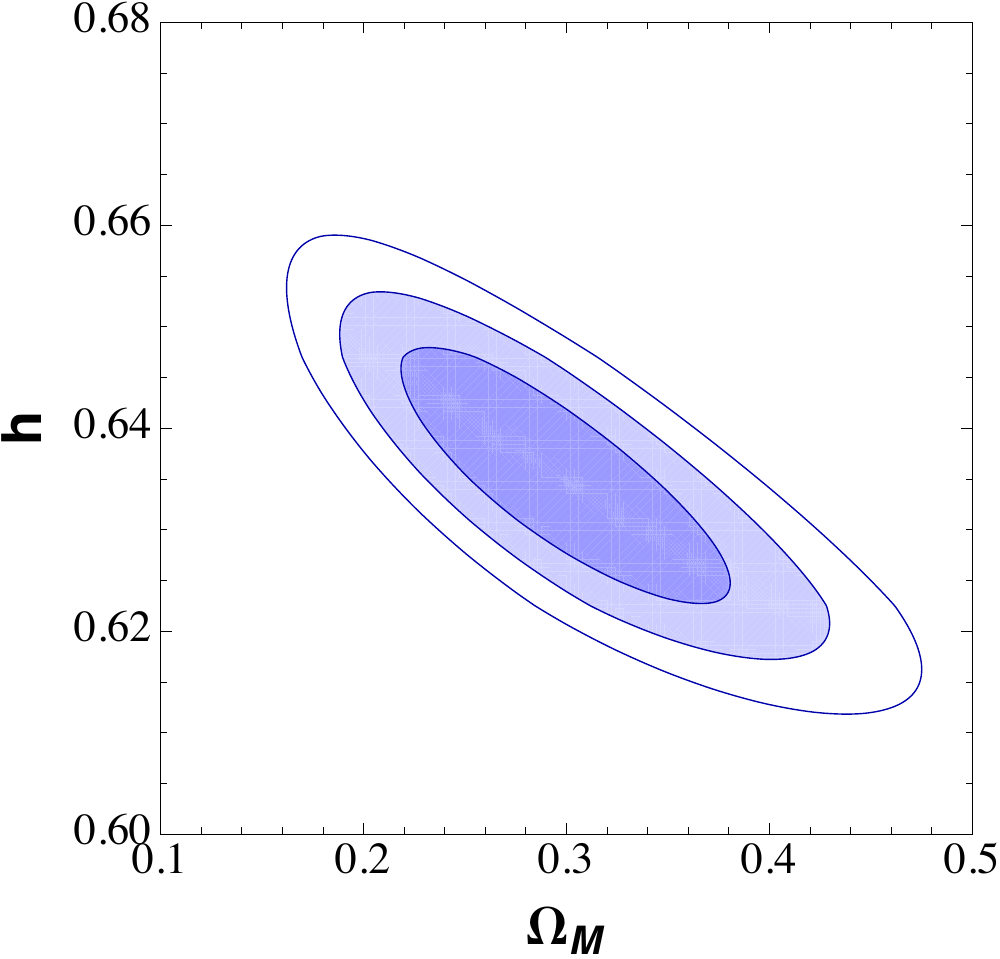} \hspace{.7cm} \includegraphics[height=5cm]{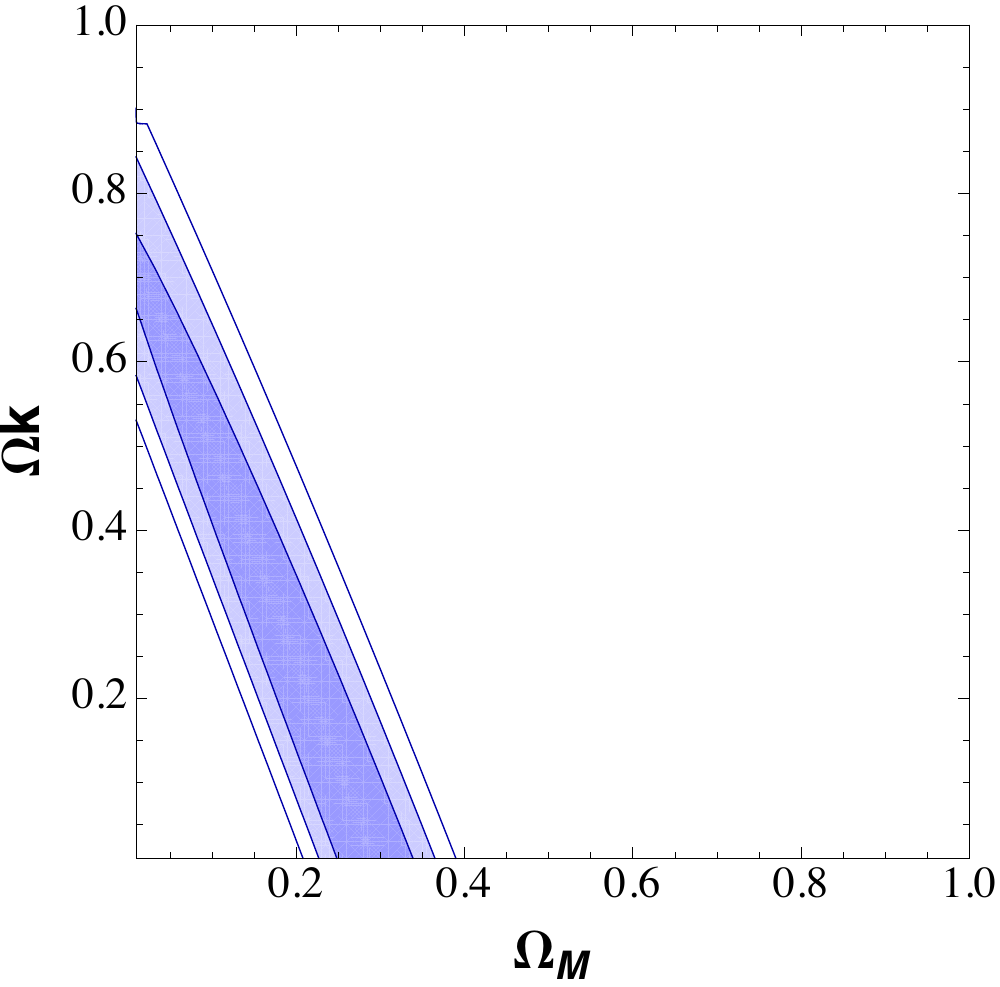} \hspace{.7cm} \includegraphics[height=4.9cm]{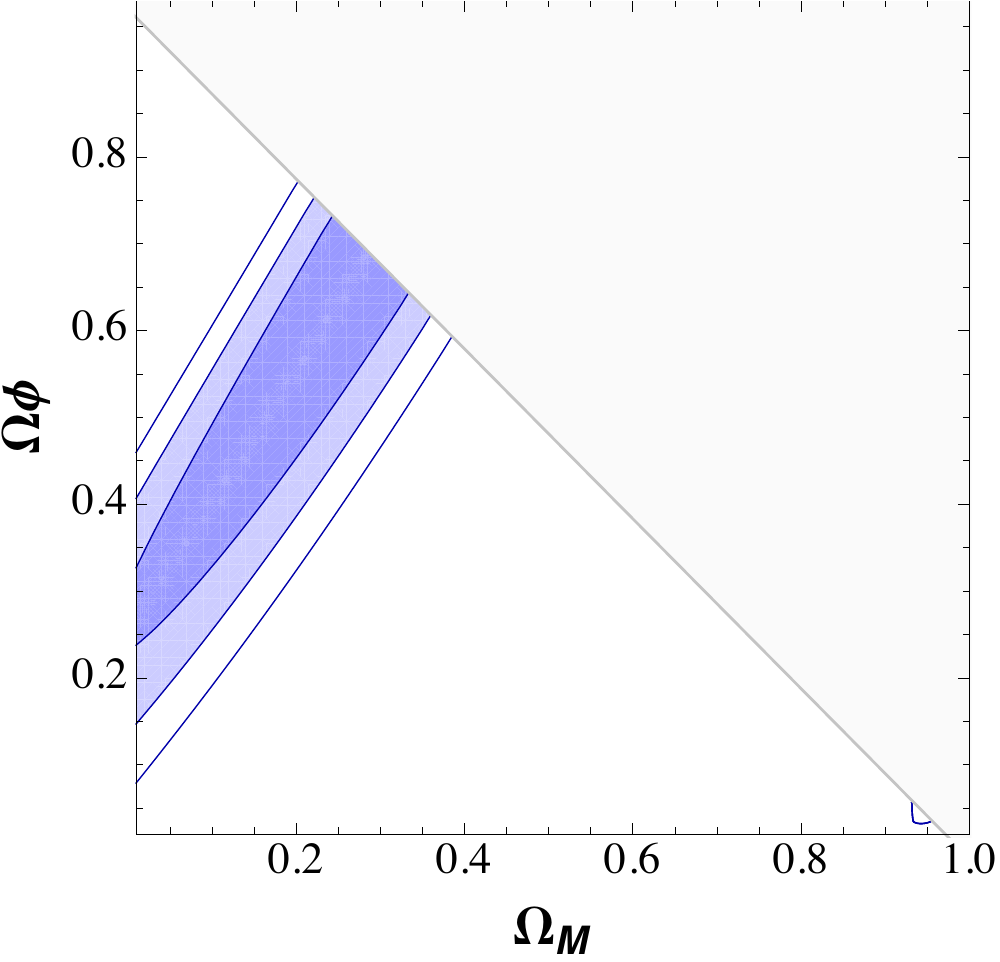}
\vspace{-0.3cm}\caption{\small{Countours of $\chi^2$ (1$\sigma$, 2$\sigma$ and 3$\sigma$ confidence regions) for the $\phi$CDM model with RTKO metric and Ratra-Peebles potential parameter $\alpha = 1$.}}
\label{fig3}
\end{figure}

\begin{figure}[t]
\includegraphics[height=5cm]{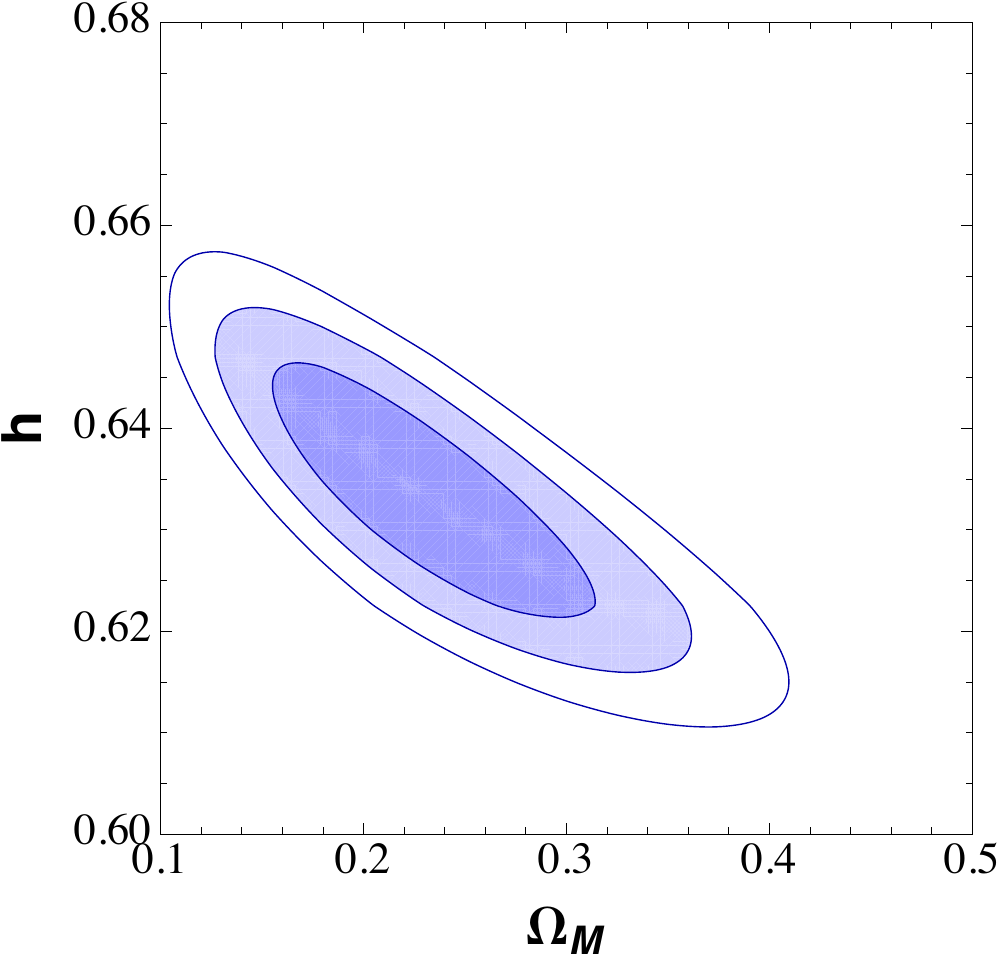}\hspace{.7cm}\includegraphics[height=5cm]{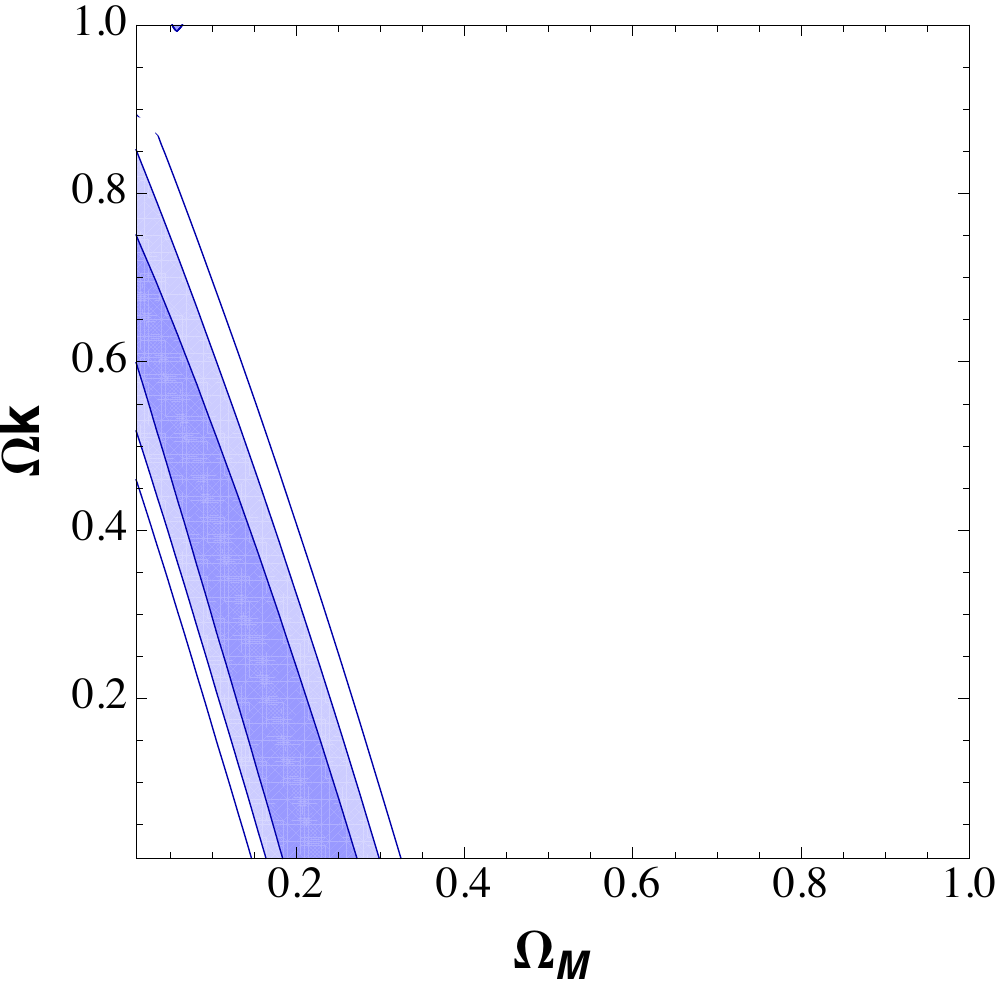}\hspace{.7cm}\includegraphics[height=5cm]{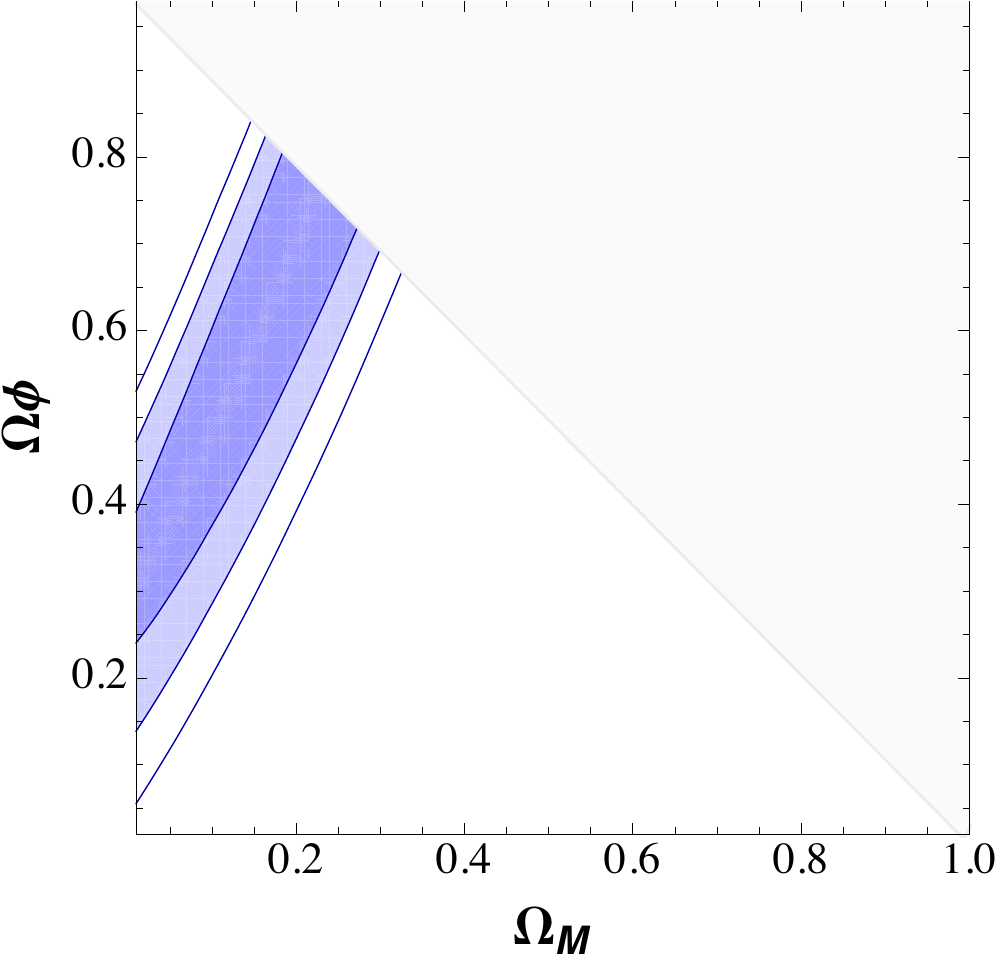}
\vspace{-0.3cm}\caption{\small{Countours of $\chi^2$ (1$\sigma$, 2$\sigma$ and 3$\sigma$ confidence regions) for the $\phi$CDM model with RTKO metric and Ratra-Peebles potential parameter $\alpha = 2$.}}
\label{fig4}
\end{figure}

We can now perform a Bayesian analysis with some SNe Ia data sample. We will analyze here the SDSS compilation \cite{SDSS}, with data of $288$ SNe Ia calibrated with the MLCS2k2 light-curve fitter \cite{mlcs2k2}, distributed in the redshift interval $[0.02, 1.55]$. This specific fitter was chosen in order to perform a model-independent test, since its calibration is made using low redshift SNe Ia \cite{tension}. For comparison, we will consider four different models, the $\Lambda$CDM and $\phi$CDM models in both the spatially flat FLRW and RTKO spacetimes. In the FLRW case the free parameters to be adjusted are $\{\Omega_M, h, M\}$\footnote{Here, $h$ is the adimensional Hubble constant defined by $H_0 = 100 h$ (km/s)/Mpc.}. For the RTKO metric they are $\{\Omega_M,\Omega_k, h, M\}$. In this last case, the analysis is made with the angular average of the luminosity distance for a given redshift. Its importante to notice that the $M$ parameter was fixed numerically and will not be analyzed here. In figures \ref{fig3} and \ref{fig4} we present confidence regions for different pairs of parameters. A summary of our results is presented in Table I. The overall effect of the quintessence scalar field, compared to the constant $\Lambda$ case, is a reduction of about $0.1$ in the present matter relative density, requiring a higher dark energy density. For the quintessence equation-of-state parameter we obtain the best-fit values $w_0 = -0.65$ for $\alpha = 1$, and $w_0=-0.5$ for $\alpha = 2$. These results are correlated with data in table I: the quintessence field produces an anti-gravitational effect less intense than the cosmological constant, requiring in this way a higher dark energy density. Table II shows a comparison of the models only in the RTKO case. If we compare the RTKO models, we can see a slightly reduction of the spatial curvature in the quintessence case. This is also because we need a higher quintessence energy density for obtaining the luminosity distances required by SNe Ia observations. A comparison of the curves for the three best-fit models, $\Lambda$CDM and $\phi$CDM with $\alpha = 1$ and $\alpha = 2$, is shown in figure 5 (left). If we analyze the anisotropy, we find that the present supernovae data cannot, in the realm of RTKO metrics, discriminate the existence of a preferred axis. Even if we set our model to the maximum anisotropy permitted by the data, $\Omega_k =0.6$, the result is virtually indistinguishable from the minimal curvature case, $\Omega_k=0.01$. To scrutinize this we have also fixed $\Omega_M=0.3$ and analyzed the $\Omega_k$ dependence of the relative deviation $\frac{\Delta \mu}{\mu} \equiv \frac{\mu(\theta = \pi/2)- \mu(\theta=0)}{\mu(\theta=0)}$. The distinction between the curves is pallid even for the highest permissible curvature (in 2$\sigma$), $\Omega_k =0.6$. It can be seen that, even for redshifts $z \sim 2$, the spatial curvature would produce a deviation in distance modulus, as seen in figure 5 (right) at most of the order of $10^{-5}$. For the best-fit value $\Omega_k = 0.01$, it would be necessary an increase in precision higher than $10^{-7}$ to distinguish a preferred axis within SNIa data, nevertheless, a small anisotropy is not necessarily ruled out. On the other hand, we see that quintessence models guarantee a very good agreement with SNe Ia observations, even on an anisotropic metric, which entails quintessence as a good dark energy candidate.

\begin{figure}[t,h]\label{fig 5}
\includegraphics[height=3.9cm]{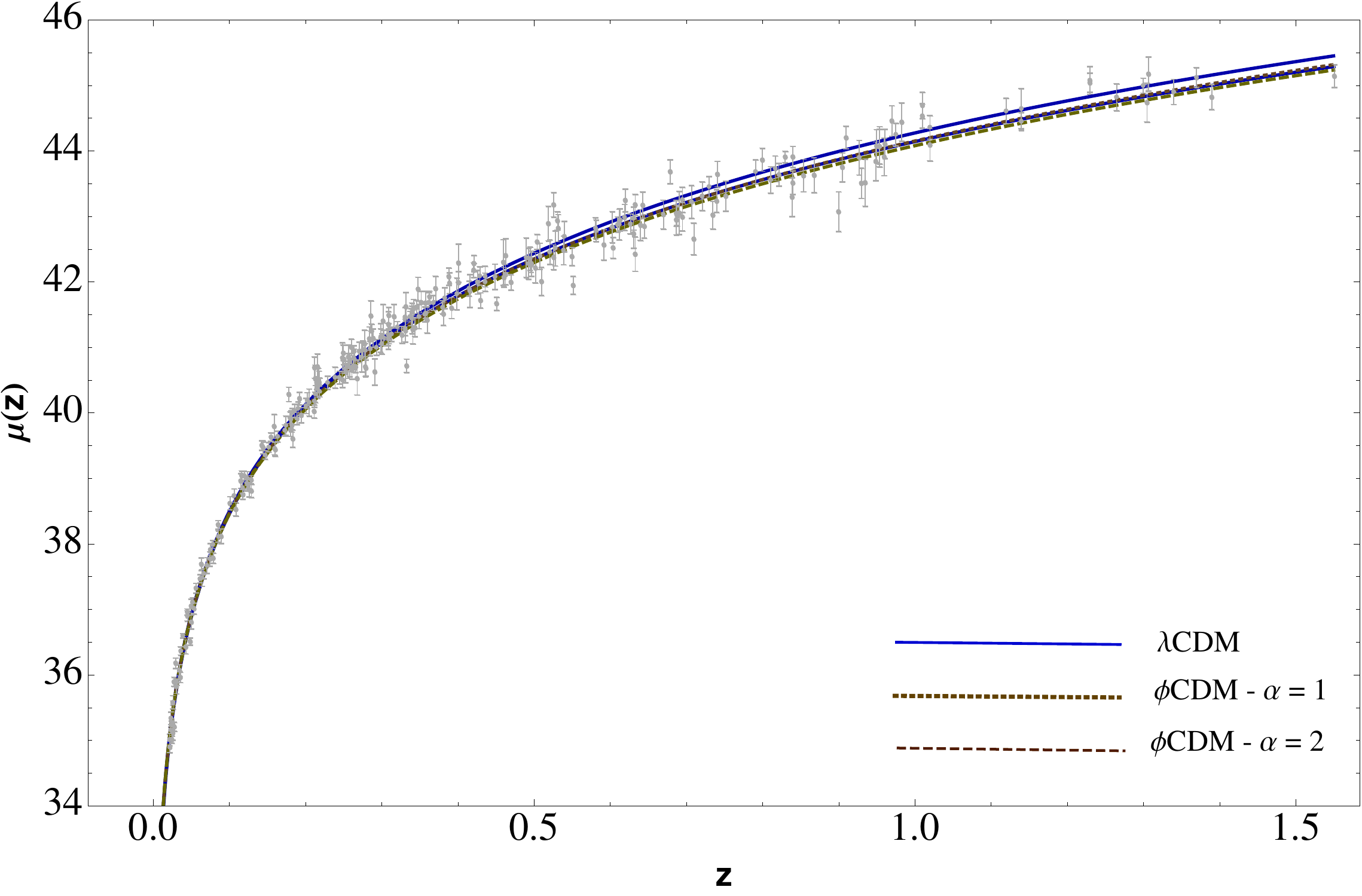}\hspace{.02cm}\includegraphics[height=3.9cm]{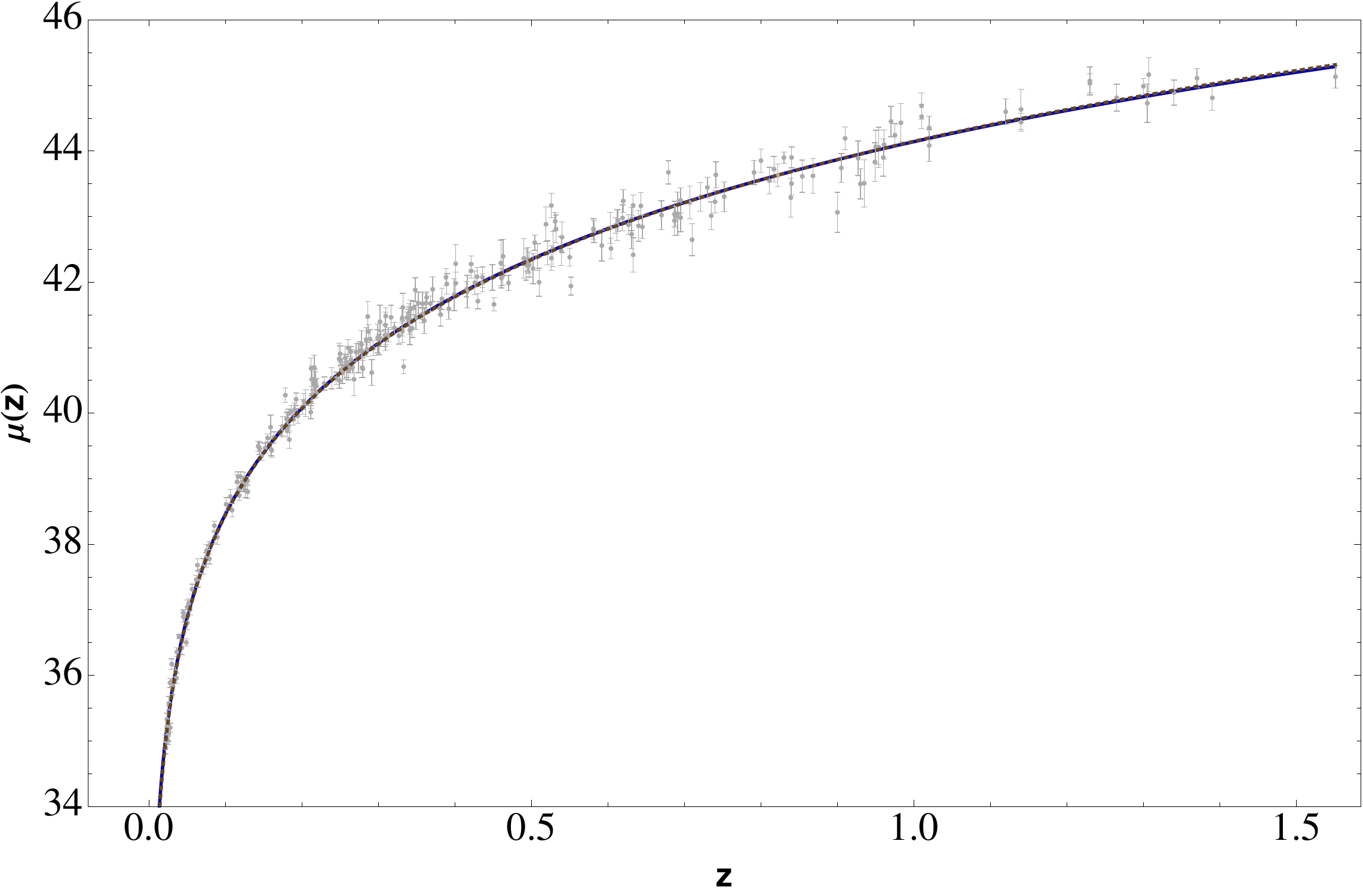}\hspace{.02cm}\includegraphics[width=5.9cm,height=3.8cm]{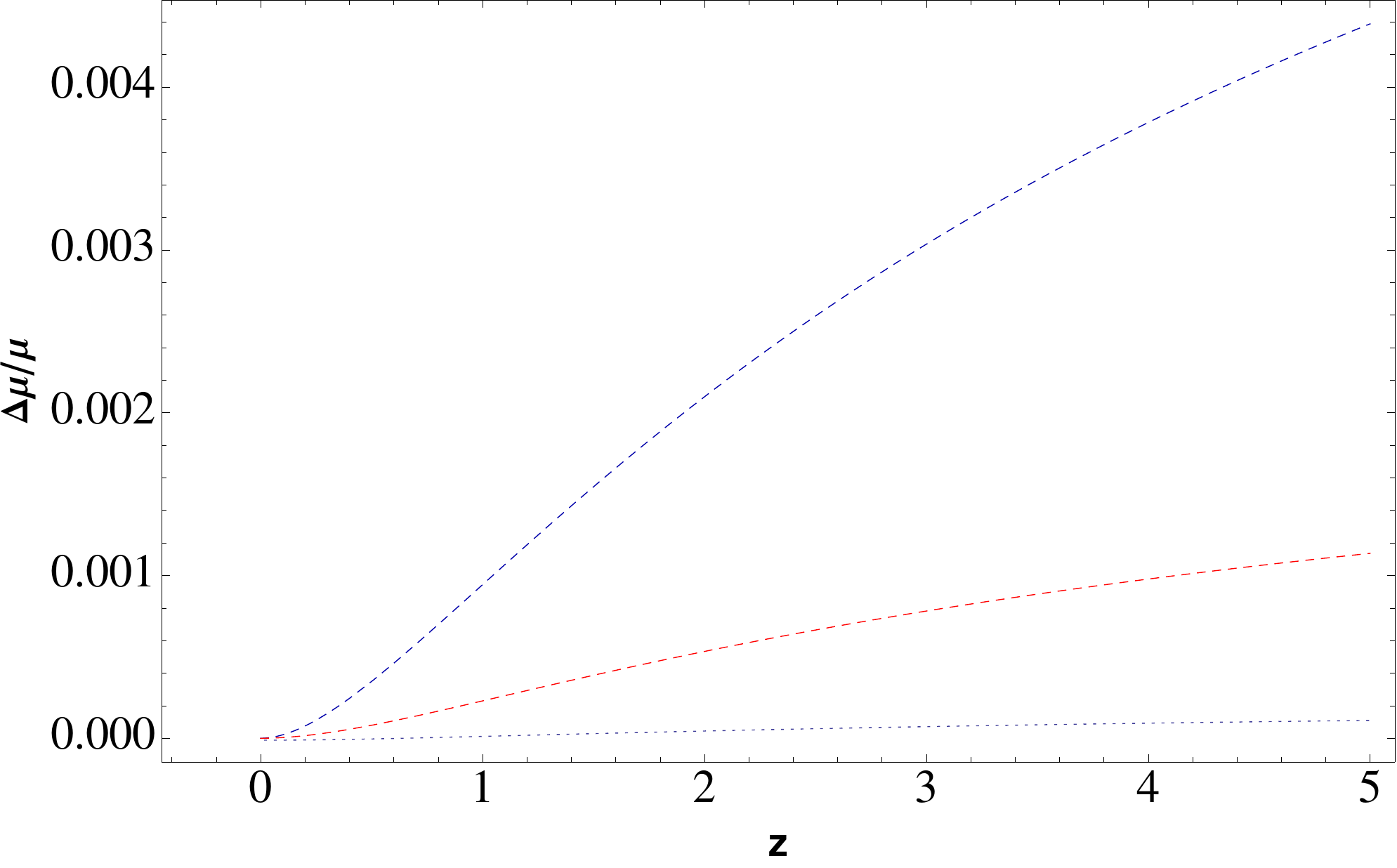}
\vspace{-0.2cm}\caption{\small{Modulus distance with best-fit $\Omega_{k}=0.01$, for the analyzed models (left). The same, superimposed curves for $\theta=\pi/2$, the $\theta$-average, and $\theta=0$ (center). Relative deviations between the maximal ($\theta=\pi/2$) and minimal ($\theta=0$) distance modulus (right).}}
\end{figure}

\begin{table}[h]\label{table1}
\begin{center}
\begin{tabular}{c|c|c|c|c|c} \hline Model & $\Omega_{M}$ & $\Omega_{k}$ & $\Omega_{\phi}$ & $q_0$ & $w_0$ \\ \hline $\Lambda CDM_{FLRW}$ & $0.40_{-0.10}^{+0.10}$  & $\equiv 0$           & $0.60$ & $-0.39$  &  $-1.00$ \\ $\phi CDM_{FLRW}$ \ ($\alpha=1$) & $0.30_{-0.11}^{+0.14}$  & $\equiv 0$           & $0.70$ & $-0.29$  &  $-0.65$ \\ $\phi CDM_{FLRW}$ \ ($\alpha=2$) & $0.23_{-0.15}^{+0.15}$ & $\equiv 0$           & $0.77$ & $-0.35$  &  $-0.50$ \\$\phi CDM_{RTKO}$ \ ($\alpha=1$) & $0.29_{-0.11}^{+0.14}$ & $0.01_{-0.8}^{+0.8}$ & $0.70$ & $-0.29$  &  $-0.65$ \\$\phi CDM_{RTKO}$ \ ($\alpha=2$) & $0.23_{-0.15}^{+0.15}$ & $0.01_{-0.8}^{+0.8}$ & $0.76$ & $-0.25$  &  $-0.50$ \\ \hline \end{tabular}
\end{center} \vspace{-0.2cm}
\caption{Best fit for some analyzed models. $q_0$ and $\omega_0$ refer, respectively, to current deceleration and quintessence equation-of-state parameters.}
\end{table}

\begin{table}[h!]\label{table2}
\begin{center}
\begin{tabular}{cc|cc|cc}\hline $\Lambda$CDM  & & $\phi$CDM & $(\alpha = 1)$& $\phi$CDM &  $(\alpha = 2)$ \\$\Omega_M$ & $\Omega_k$ & $\Omega_M$ & $\Omega_k$ & $\Omega_M$ & $\Omega_k$ \\\hline$0.38_{-0.4}^{+0.1}$ & $0.045_{-0.6}^{+0.6}$ & $0.29_{-0.11}^{+0.14}$ & $0.01_{-0.4}^{+0.4}$ & $0.23_{-0.15}^{+0.15}$ & $0.01_{-0.4}^{+0.4}$ \\\hline\end{tabular}
\end{center}\vspace{-0.2cm}
\caption{Best-fit parameters for the RTKO metric with $2\sigma$ confidence level}
\end{table}

\section{Conclusions}

We have studied a quintessence model with a Bianchi type III metric. Although the particular case considered here has zero rotation, there is a preferred axis in this spacetime, which is dependable on the metric curvature. We have analysed the effects of anisotropic curvature and the dynamics of a quintessence model and compared to the $\Lambda$CDM model in both isotropic and anisotropic RTKO metric. An anisotropic, time-independent field $\psi$ is responsible for the space anisotopy, and another dynamical quintessence field $\phi$ acts as a dark energy component with negative pressure. The chosen Ratra-Peebles potential has led to a cosmological dynamics that is very suitable for the description of the cosmos, because its tracker behavior does not require a very specific adjustment of the initial conditions on the scalar field energy density . This alleviates the fine-tuning problem present in the standard $\Lambda$CDM model. We have tested the models against the SDSS compilation of 288 SNe Ia luminosity-distance data. Our analysis has led to an observational fit virtually indistinguishable from the standard model with constant $\Lambda$. The best fits in the quintessence cases correspond to lower values of curvature and matter densities compared to the standard model,  and to a higher $\Omega_{\phi}$. We have seen firstly that, the resulting quintessence model in a curved anisotropic space-time has the required properties to be a suitable dark energy candidate and secondly that, even for the highest redshifts available for SNIa data, the current precision of data does not allow to detect a cosmic anisotropy within that analysis.

\section*{Acknowledgements} The authors are thankful to CNPq (Brazil) for the grants under which this work was carried out, to Roberto Menezes for useful discussions, and to Rita Novaes for a revision.

{}\end{document}